# DO MUSIC GENERATION MODELS ENCODE MUSIC THEORY?

**Megan Wei**[1*]    **Michael Freeman**[1*]    **Chris Donahue**[2]    **Chen Sun**[1]

[1] Brown University    [2] Carnegie Mellon University

`meganwei@brown.edu, michael_freeman@alumni.brown.edu`

## ABSTRACT

Music foundation models possess impressive music generation capabilities. When people compose music, they may infuse their understanding of music into their work, by using notes and intervals to craft melodies, chords to build progressions, and tempo to create a rhythmic feel. To what extent is this true of music generation models? More specifically, are fundamental Western music theory concepts observable within the "inner workings" of these models? Recent work proposed leveraging latent audio representations from music generation models towards music information retrieval tasks (e.g. genre classification, emotion recognition), which suggests that high-level musical characteristics are encoded within these models. However, probing individual music theory concepts (e.g. tempo, pitch class, chord quality) remains under-explored. Thus, we introduce **SynTheory**, a synthetic MIDI and audio music theory dataset, consisting of tempos, time signatures, notes, intervals, scales, chords, and chord progressions concepts. We then propose a framework to probe for these music theory concepts in music foundation models (Jukebox and MusicGen) and assess how strongly they encode these concepts within their internal representations. Our findings suggest that music theory concepts are discernible within foundation models and that the degree to which they are detectable varies by model size and layer.

## 1. INTRODUCTION

State-of-the-art text-to-music generative models [1–3] exhibit impressive generative capabilities. Past work suggests that internal representations of audio extracted from music generative models encode information relating to high-level concepts (e.g. genre, instruments, or emotion) [4–7]. However, it remains unclear if they also capture underlying symbolic music concepts (e.g. tempo or chord progressions) [8].

We aim to investigate whether and to what extent state-of-the-art music generation models encode music theory concepts in their internal representations. Confirming this could enable the creative alteration of these concepts, providing artists with new methods towards more detailed and lower-level control [9] (e.g. changing the key of a song or editing a particular chord in a chord progression). Furthermore, by benchmarking these foundation models, we identify potential avenues for improvement towards stronger concept encoding. Our approach builds upon work in probing and editing concepts in language models, which have shown promise in identifying emergent representations in autoregressive models and editing factual knowledge [9–12]. While previous work probed music generation models for high-level concepts, such as emotion, genre, and tagging [4–7], we focus on uncovering representations of more granular music theory concepts. Although existing datasets such as HookTheory [13] contain rich music theory annotations, their association with copyrighted music potentially complicates their use.

---

*: Equal contribution.

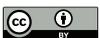 

Our first contribution is a framework that generates diagnostic datasets for probing music theory concepts in music generation models, allowing programmatic control over concept selection while mitigating the presence of potential distractors. Our synthetic music theory dataset, **SynTheory**, consists of seven music concepts based on Western music theory: tempo, time signatures, notes, intervals, scales, chords, and chord progressions. SynTheory serves as a customizable, copyright-free, and scalable approach towards generating diagnostic music clips for probing music generative models trained on real-world data.

Our second contribution is the analysis of two state-of-the-art music generative models Jukebox [3] and MusicGen [1] with our SynTheory benchmark. We extract representations for the concepts defined in SynTheory from MusicGen and Jukebox and assess whether these models encode meaningful representations of these concepts. To analyze the internal representations of these models, we train probing classifiers [14] based on ground truth music theory concept labels. A higher classification accuracy implies that these models learn internal representations that "understand" music theory concepts, which can be decoded by a multi-layer perceptron (MLP) or a linear model.

Our results show that music foundation models encode meaningful representations of music theory concepts. These representations vary across different sections of the model (audio codecs, decoder LMs), different layers within the decoder LMs, and different model sizes. Furthermore, the nature of the concepts, from time-varying (e.g. chord progressions) to stationary (e.g. notes, chords), influence the performance of these models across these tasks. We hope our insights on probing music foundation models, along with the synthetic music data generation framework, encourage and facilitate future endeavors on symbolic controllability in music generative models.

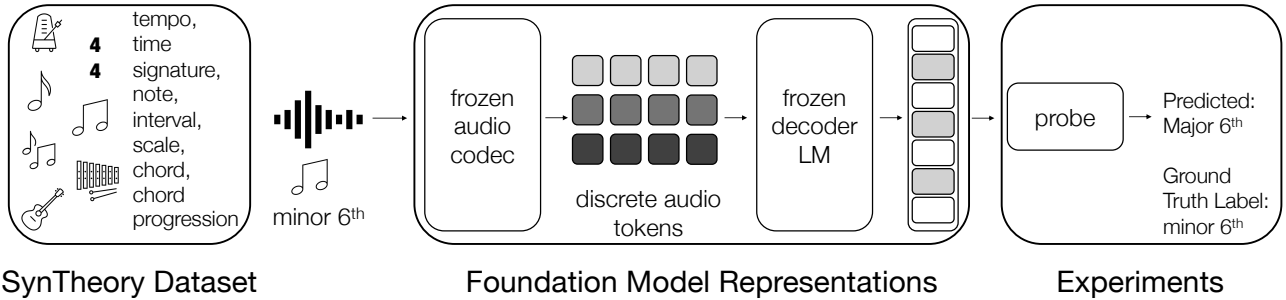

**Figure 1**. Overview of our SynTheory benchmark and our Jukebox and MusicGen probing setup. Our SynTheory benchmark consists of **Rhythmic** (tempos and time signatures) and **Tonal** (notes, intervals, scales, chords, and chord progressions) concepts. We assess whether music foundation models (Jukebox and MusicGen) encode these music theory concepts within their internal representations. For each task from the SynTheory dataset, we extract representations from the music foundation model. We pass an audio input, embodying the concept (e.g. Perfect 4th), into a pretrained foundation model. The audio codec tokenizes the audio into discrete audio tokens. Then, it passes these tokens into a decoder language model. From there, we extract the representations. We then train a probe classifier (linear or two-layer MLP) on these representations to predict the corresponding class (e.g. pitch class, intervals, or chords) for each SynTheory concept.

For reproducibility, we upload our SynTheory dataset to Hugging Face [1], release the code for dataset generation, embedding extraction, probing, and evaluation on GitHub [2], and showcase audio samples on our website [3].

## 2. RELATED WORK

The success of large language models (LLMs) [15–18] has sparked new research on probing their internal representations to measure their understanding of linguistic concepts [19, 20] and world knowledge [11, 12, 21] as well as editing the encoded knowledge to make LLMs more faithful to factual knowledge [9, 10]. Studies have shown that LLMs can encode grounded representations on color [22], direction [23], and auditory representations [24]. Thus, we investigate if music generative models, which often share similar model architectures and training objectives as LLMs, are able to encode abstract concepts from high-level music tags (e.g. genre, emotion) to fine-grained music theory knowledge (e.g. tempo, chords).

Recent work has shown promise towards uncovering conceptual representations from probing audio and music generative models and leveraging different music foundation model architectures towards music understanding tasks. Castellon and Donahue et al. [4] propose using representations from language models trained on codified audio towards downstream MIR tasks as a better alternative to conventional tagging models. The authors train probing classifiers on Jukebox representations on music tagging, genre identification, key identification, and emotion recognition tasks. These results demonstrate the effectiveness of using internal model representations for downstream MIR tasks. Koo et al. [7] focus primarily on probing MusicGen's attention heads in instrument recognition tasks, benchmarking against the tasks highlighted in [4], and propose leveraging these representations for inference-time control. Other works [5,6] assess the impact of model architectures and self-supervised approaches towards music understanding tasks.

However, prior work primarily uses real-world data, which is often concept-entangled and potentially subject to copyright concerns. For example, some of these works use Giantsteps-MTG and Giantsteps [25], which are datasets of primarily electronic dance music with tempo and key annotations, obtained from Beatport. Won et al. [5] use HookTheory for chord recognition, where they focus on major and minor chord identification for each pitch class. The authors also use Harmonix Set [26] and GTZAN [27] for beat and downbeat detection. In the language modality, the authors of ChatMusician [28] produce a multi-choice question answering dataset, MusicTheoryBench, with expert annotation from a professional college music teacher. MusicTheoryBench aims to assess the music understanding capabilities of LLMs through natural language alone. To the best of our knowledge, there is a lack of music theory probing benchmarks in the audio domain that are labeled in detail, copyright-free, and scalable, prior to our proposed SynTheory.

## 3. SYNTHEORY: SYNTHETIC DATASET OF MUSIC THEORY CONCEPTS

We design seven datasets to capture isolated music theory concepts – similar to synthetic audio for ear training. Musicians may "train their ear" to recognize music concepts like intervals or chord quality in an isolated setting before advancing to the harder, more entangled case that arises in non-pedagogical music. Assessing concept recognition through isolated concepts mitigates the possibility that one intuits or guesses the answer from its context. Literature on instrument-specific absolute pitch in humans corroborates the notion that timbral information may be exploited in identifying a different concept like pitch class [29]. As such, our dataset is designed to remove or reduce features that may correlate with a concept, but are not strictly necessary for identifying it. Our intent is a more pointed assessment towards theoretical concepts as abstract ideas rather than as acoustically realized audio. A more practical motivation for this work is that extracting such low-level, isolated concepts from existing datasets may require non-trivial engineering or domain expert labor. It may even be impossible to disentangle all overlapping concepts. Music stem isolation and concept isolation are distinct; an isolated instrument in a multi-track recording may still exhibit several, intricately intertwined theory concepts. It is not clear how to "unmix" such concepts once they are blended.

---
[1] https://huggingface.co/datasets/meganwei/syntheory
[2] https://github.com/brown-palm/syntheory
[3] https://brown-palm.github.io/music-theory

Instead of attempting to disentangle several concepts from existing audio, SynTheory implements this "ear training" quiz setting by explicitly producing individual concepts. Each of the seven datasets ablates a single musical feature while fixing all others, thereby isolating it to a degree not typically found in recorded music. These ablated concepts consist of tempo, time signatures, notes, intervals, scales, chords, and chord progressions. We adopt isolation as a design choice to mitigate context that may be exploited in deep learning models as "shortcuts", i.e. heuristics that correlate with concepts most of the time but do not truly encode the concept.

Using this music theory concept-driven synthesis design, we construct label-balanced and copyright-free data. The synthetic approach avoids annotation errors present in other contemporary MIR datasets. For example, the HookTheory data processing step for SheetSage [13] required ad-hoc time-alignment of the expert annotations. In the released SheetSage dataset, $17,980/26,175$ (68.7%) samples required more precise time alignment. While our synthetic data is no substitute for real music data, to our knowledge, no other dataset so strictly isolates each concept to this level of granularity.

SynTheory contains two categories: tonal and rhythmic. We make this distinction for stronger concept isolation; we wish to keep the rhythm samples tonally consistent and the tonal samples rhythmically consistent. For each **tonal** dataset, we voice the same MIDI data through 92 distinct instruments. The selection of instrument voices is fixed, making the distribution of timbres sufficiently diverse but also class-balanced. Each instrument corresponds to one of the 128 canonical MIDI program codes and is voiced through the `TimGM6mb.sf2` [30] soundfont. A MIDI "program" is a specific instrument preset. The canonical program set includes many instruments, e.g. "Acoustic Grand Piano", "Flute", etc. We exclude programs that are polyphonic, sound effects (e.g. "Bird Tweet", "Gun Shot"), and highly articulate. A highly articulate program has some unchangeable characteristic (e.g. pitch bending) that destabilizes its pitch. For each **rhythmic** dataset, we define five metronome-like timbral settings. Each setting uses one of the distinct instruments: "Woodblock Light", "Woodblock Dark", "Taiko", "Synth Drum", and the MIDI drum-kit, following the voicing done in Sheetsage [13]. Each setting produces a distinct sound on the upbeat and the downbeats, which defines the time signature concept.

| Concept | Total Samples |
|---|---|
| Tempo | 4,025 |
| Time Signatures | 1,200 |
| Notes | 9,936 [4] |
| Intervals | 39,744 |
| Scales | 15,456 |
| Chords | 13,248 |
| Chord Progressions | 20,976 |

**Table 1**. **SynTheory** contains seven synthetic datasets, each of which captures an isolated music theory concept. We present an overview of these datasets and their sizes.

One can extend or alter these configurations using the SynTheory codebase. We provide a framework that enables declarative and programmatic MIDI construction in musical semantics, audio export in any soundfont, and dataset construction for use in our framework.

### 3.1 SynTheory-Rhythmic

*3.1.1 Tempo*

We voice integer tempi from 50 to 210 BPM (beats per minute) inclusive in $\frac{4}{4}$ time. To ensure diverse start times, we produce five random offset times per sample. There are $(5 \text{ CLICK SETTING} \cdot 161 \text{ TEMPO} \cdot 5 \text{ OFFSET}) = 4,025$ samples in total.

*3.1.2 Time Signature*

We voice the following time signatures: $\frac{2}{2}, \frac{2}{4}, \frac{3}{4}, \frac{3}{8}, \frac{4}{4}, \frac{6}{8}, \frac{9}{8}$, and $\frac{12}{8}$ at a fixed tempo of 120 BPM. To add acoustic variation, we add three levels of reverb from completely dry to spacious. We find empirically that this acoustic perturbation increases the difficulty of the probing task. Like the *Tempo* dataset, we produce ten random offset times per sample. There are $(8 \text{ TIME SIGNATURE} \cdot 3 \text{ REVERB LEVEL} \cdot 5 \text{ CLICK SETTING} \cdot 10 \text{ OFFSET}) = 1,200$ samples.

### 3.2 SynTheory-Tonal

*3.2.1 Notes*

We voice all twelve Western temperament pitch classes, in nine octaves, using 92 instruments. The note is played in quarter notes at a tempo of 120 BPM, with no distinction between the upbeat or downbeat. There are $(12 \text{ PITCH CLASS} \cdot 9 \text{ OCTAVE} \cdot 92 \text{ INSTRUMENT}) = 9,936$ configurations. However, there are only 9,848 distinct samples because 88 configurations at extreme registers are unvoiceable in our soundfont. These silent samples are listed for completeness in our Hugging Face dataset.

---

[4] There are 9,936 distinct note configurations, but our dataset contains 9,848 non-silent samples. With a more complete soundfont, all 9,936 configurations are realizable to audio.

### 3.2.2 Intervals

We vary the root note, number of half-steps, instrument, and play style (unison, up, and down). To retain consistent rhythm, the up and down styles repeat four times throughout the sample while the unison play style repeats eight times. There are (12 PITCH CLASS · 12 HALF-STEP · 92 INSTRUMENT · 3 PLAY STYLE) = 39,744 samples.

### 3.2.3 Scales

We voice seven Western music modes (Ionian, Dorian, Phrygian, Lydian, Mixolydian, Aeolian, and Locrian) in all root notes, in 92 instruments, and in two play styles (ascending or descending). The register is constant; we select root notes close to middle C. There are (7 MODE · 12 ROOT NOTE · 92 INSTRUMENT · 2 PLAY STYLE) = 15,456 samples.

### 3.2.4 Chords

We voice triads of all twelve root notes, four chord qualities (major, minor, augmented, and diminished), 92 instruments, and three inversions (root position, first inversion, and second inversion). The chord is struck at each quarter note at 120 BPM. Like the *Scales* dataset, we fix the register close to middle C. There are (12 ROOT NOTE · 4 CHORD QUALITY · 92 INSTRUMENT · 3 INVERSION) = 13,248 samples.

### 3.2.5 Chord Progressions

We select 19 four-chord progressions, with ten in the major mode and nine in the natural minor mode:

- Major: (I–IV–V–I), (I–IV–vi–V), (I–V–vi–IV), (I–vi–IV–V), (ii–V–I–vi), (IV–I–V–vi), (IV–V–iii–vi), (V–IV–I–V), (V–vi–IV–I), (vi–IV–I–V)

- Natural Minor: (i–ii°–v–i), (i–III–iv–i), (i–iv–v–i), (i–VI–III–VII), (i–VI–VII–i), (i–VI–VII–III), (i–VII–VI–IV), (iv–VII–i–i), (VII–vi–VII–i)

We vary the root note of the key and instrument. Each chord is played at each quarter notes at 120 BPM. There are (19 PROGRESSION · 12 KEY ROOT · 92 INSTRUMENT) = 20,976 samples.

## 4. EXPERIMENTS

We describe our approach to assess how well the internal representations of music generative models (MusicGen and Jukebox) and handcrafted audio features (mel spectrograms, MFCC, and chroma) encode music theory concepts.

### 4.1 Evaluation

A "probe" is a simple classifier, often a linear model, trained on the activations of a neural network [14]. Accurate performance of such classifiers suggests that information relevant to the class exists in the latent representations within the network. As such, probes may be used as a proxy for measuring a model's "understanding" or encoding of abstract concepts. Motivated by the use of probes to discover linguistic structure and semantics in NLP [31] and more recently in MIR [4], we use probes to assess whether music theory concepts are discernable in the internal representations of foundation models.

We adopt the same probing paradigm as [4] and frame concept understanding as multiclass classification for discrete concepts (notes, intervals, scales, chords, chord progressions, and time signatures) and regression for continuous concepts (tempo). We train linear and two-layer MLP probes on the embeddings of the internal representations of Jukebox and MusicGen and the handcrafted features.

Each probe is trained independently for its corresponding concept task. That is, the probe trained to identify notes from Jukebox embeddings will not be used to identify intervals, for example.

For the classification tasks, we measure the accuracy of our trained probes on the following SynTheory tasks:

- Notes (12): C, C#, D, D#, E, F, F#, G, G#, A, A#, and B
- Intervals (12): minor 2nd, Major 2nd, minor 3rd, Major 3rd, Perfect 4th, Tritone, Perfect 5th, minor 6th, Major 6th, minor 7th, Major 7th, and Perfect octave
- Scales (7): Ionian, Dorian, Phrygian, Lydian, Mixolydian, Aeolian, and Locrian
- Chords (4): Major, Minor, Diminished, and Augmented
- Chord Progressions (19): (I–IV–V–I), (I–IV–vi–V), (I–V–vi–IV), (I–vi–IV–V), (ii–V–I–vi), (IV–I–V–vi), (IV–V–iii–vi), (V–IV–I–V), (V–vi–IV–I), (vi–IV–I–V), (i–ii°–v–i), (i–III–iv–i), (i–iv–v–i), (i–VI–III–VII), (i–VI–VII–i), (i–VI–VII–III), (i–VII–VI–IV), (iv–VII–i–i), and (VII–vi–VII–i)
- Time Signatures (8): $\frac{2}{2}$, $\frac{2}{4}$, $\frac{3}{4}$, $\frac{3}{8}$, $\frac{4}{4}$, $\frac{6}{8}$, $\frac{9}{8}$, and $\frac{12}{8}$

These tasks are trained on a 70% train, 15% test, and 15% validation split, using the Adam optimizer and Cross Entropy loss.

For the *Tempo* dataset, we train a regression probe, over the 161 tempo values. To increase complexity in the probing task and test generalization to unseen BPMs, the training set consists of the middle 70% of the BPMs. The test and validation sets consist of the top 15% BPMs and the bottom 15% BPMs, randomly shuffled and split in half. We use MSE loss and report the $R^2$ score.

To select the best performing probe for each concept, we perform a grid search across various hyperparameters for the MusicGen audio codec, mel spectrogram, MFCC, chroma, and aggregate handcrafted features, following [4]:

- Data Normalization: {True, False}
- Model Type: {Linear, two-layer MLP with 512 hidden units and ReLU activation}
- Batch Size: {64, 256}
- Learning Rate: {$10^{-5}$, $10^{-4}$, $10^{-3}$}
- Dropout: {0.25, 0.5, 0.75}
- L2 Weight Decay: {off, $10^{-4}$, $10^{-3}$}

For the decoder language models (MusicGen small, medium, and large and Jukebox), we use a fixed set of hyperparameters and select the probe with the best performing layer for each concept, in the interest of computational efficiency:

- Data Normalization: True
- Model Type: two-layer MLP with 512 hidden units and ReLU activation
- Batch Size: 64
- Learning Rate: $10^{-3}$
- Dropout: 0.5
- L2 Weight Decay: off

We selected these hyperparameters from the best overall performing probe by fixing a layer in the decoder LMs and performing a hyperparameter search, following the sweep approach outlined in [4].

### 4.2 Model representations

We extract representations from two text-to-music foundation models, Jukebox [3] and MusicGen [1] and benchmark them against three handcrafted, spectral features following [4]: mel spectrograms, mel-frequency cepstral coefficients (MFCC), and constant-Q chromagrams (chroma). These handcrafted features are common in traditional methods of MIR and are a more interpretable baseline against the embeddings of the pre-trained music generative models.

Jukebox consists of a VQ-VAE model that codifies audio waveforms into discrete codes at a lower sample rate and a language model that generates codified audio with a transformer decoder. We trim each audio sample to four seconds and convert it to mono. Using Jukemirlib [4], we pass the audio through the frozen audio encoder and through the decoder language model. Due to resource constraints, we downsample the activation to a target rate of half that in [4], using the Librosa FFT algorithm [32]. Then, we meanpool the representations across time to reduce the dimensionality of the embeddings, resulting a dimension of $(72, 4800)$ per sample, where 72 is the number of layers and 4800 is the dimension of the activations. To further reduce the dimensions, we adopt a layer selection process similar to [4]. We train the probe classifiers with fixed hyperparameters on the music concept tasks as described in Section 4.1. For each concept, we select the layer that results in the highest probing score. This results in a final dimension of 4800 for each Jukebox representation.

MusicGen consists of a pretrained convolutional autoencoder (EnCodec) [33], a pretrained T5 text encoder, and an acoustic transformer decoder. We resample the audio to 32 kHz (the sampling rate used in the EnCodec model) trim to four seconds, convert to mono, and pass the audio through the frozen EnCodec audio codec. We do not pass text through the text encoder, as we focus on audio representations. We then extract representations from several regions of the model: the final layer of the audio codec before residual vector quantization and the hidden states of the decoder language model. The number of decoder hidden states vary based on the model size: small (24 layers), medium (48 layers), and large (48 layers).

For our four second audio clips, the audio codec representations are of dimension $(128, 200)$, where 128 is the dimension of the activation after the final layer of the au-dio codec and 200 is the sequence length. We meanpool the values of the representations across time, resulting in a final dimension of 128 for the MusicGen audio codec.

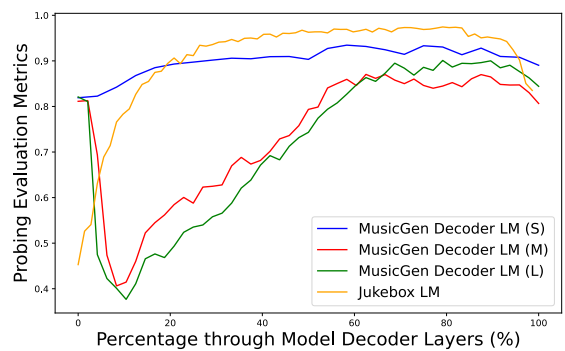

**Figure 2**. Probing evaluation metrics averaged across all SynTheory concepts over the model layers of Jukebox and MusicGen decoder models. The probing evaluation metric is $R^2$ for tempos and accuracy for the rest of the SynTheory concepts (notes, intervals, scales, chords, chord progressions, and time signatures). Features extracted from deeper layers generally perform better, with a slight drop-off near the final layers.

The decoder hidden states for the small, medium, and large MusicGen models have dimensions $(24, 200, 1024)$, $(48, 200, 1536)$, and $(48, 200, 2048)$ respectively, where the first axis corresponds to the number of layers, second corresponds to sequence length, and third corresponds to hidden size. To reduce the dimensions, we meanpool across time, resulting in dimensions of $(24, 1024)$ for MusicGen small decoder, $(48, 1536)$ for MusicGen medium decoder, and $(48, 2048)$ for MusicGen large decoder. Similar to our approach with Jukebox, we select the most optimal layer for each decoder model size based on the probing scores. After selecting the best performing layer per concept and model size, the dimensions of the representations are 1024 for MusicGen small, 1536 for MusicGen medium, and 2048 for MusicGen large.

We visualize results from probing across layers per model (MusicGen and Jukebox) averaged across concepts in Figure 2.

We extract the handcrafted features (mel spectrograms, mel-frequency cepstral coefficients, and constant-Q chromagrams) with librosa [32]. Similar to [4], we concatenate the mean and standard deviation across time of these features along with their first- and second-order discrete differences. To obtain an aggregate representation of the handcrafted features, we concatenate the mel spectrogram, chroma, and MFCC features and obtain their mean and standard deviation across time and their first- and second-order differences.

## 5. RESULTS AND DISCUSSION

Overall, we observe that music generation models do encode music theory concepts in their internal representa-tions. Jukebox performs consistently well across all SynTheory tasks. MusicGen Decoder LMs also exhibit competitive performance across our benchmark. Interestingly, we observe that MusicGen Decoder (Small) outperforms its larger counterparts. This finding contradicts traditional discussions on scaling laws and the claims in [1] that larger MusicGen models better "understand" text prompts and produce better quantitative and subjective scores. Figure 2 further highlights this finding, showing that MusicGen Decoder (Small) maintains consistently high probing scores across all layers compared to that of its larger counterparts. Meanwhile, larger MusicGen models display a steep performance drop in the initial layers, followed by a gradual increase and then a slight decline in the final layers. This pattern suggests that the smaller model may have developed a more efficient encoding of music theory concepts within its representations.

|  | Notes | Intervals | Scales | Chords | Chord Progressions | Tempos | Time Signatures | Average |
|---|---|---|---|---|---|---|---|---|
| Jukebox LM | 0.951 | 0.995 | 0.978 | 0.997 | 0.971 | 0.993 | 1.000 | 0.984 |
| MusicGen LM (S) | 0.897 | 0.995 | 0.949 | 0.990 | 0.942 | 0.969 | 0.911 | 0.950 |
| MusicGen LM (M) | 0.851 | 0.983 | 0.863 | 0.989 | 0.870 | 0.956 | 0.883 | 0.914 |
| MusicGen LM (L) | 0.866 | 0.972 | 0.905 | 0.989 | 0.901 | 0.965 | 0.905 | 0.929 |
| MusicGen Audio Codec | 0.729 | 0.965 | 0.383 | 0.879 | 0.330 | 0.947 | 0.677 | 0.701 |
| Mel Spectrogram | 0.712 | 0.995 | 0.897 | 0.988 | 0.723 | 0.785 | 0.827 | 0.847 |
| MFCC | 0.467 | 0.822 | 0.370 | 0.863 | 0.872 | 0.923 | 0.688 | 0.715 |
| Chroma | 0.954 | 0.820 | 0.989 | 0.994 | 0.869 | 0.847 | 0.672 | 0.878 |
| Aggregate Handcrafted | 0.941 | 0.997 | 0.972 | 0.992 | 0.868 | 0.947 | 0.833 | 0.936 |

**Table 2**. We report probing results on the SynTheory dataset for the Jukebox LM, MusicGen Decoder LM (Small, Medium, and Large), MusicGen Audio Codec models as well as handcrafted features (Mel Spectrogram, MFCC, Chroma, and Aggregate Handcrafted). For the tempos dataset, we report the $R^2$ score from the regression probe. For all other concepts (notes, intervals, scales, chords, chord progressions, and time signatures), we report the probing classifier accuracy. For MusicGen Audio Codec, Mel Spectrogram, MFCC, Chroma, and Aggregate Handcrafted, we report the metrics of the best performing probe for each task using the best validation performance from our hyperparameter search. For MusicGen Decoder LM (Small, Medium, and Large) and Jukebox models, we report the metrics of the best performing probe for each task using layer selection. We also report an average performance across all concepts for each model/feature.

While these models perform well on other concepts, MusicGen slightly underperforms on the notes dataset. We hypothesize this may be due to isolated notes being less prominent as intervals, scales, and chords in real-world music. This reveals the challenge in distinguishing the most fundamental building blocks of music.

Pretrained music decoder LMs generally outperform MusicGen Audio Codec representations and individual handcrafted features. MusicGen Audio Codec exhibits poorer overall performance, since these codecs were trained to reconstruct fine-grained, low-level details localized by time.

Chroma features, which encode pitch class information, perform comparably well on tonal tasks, but slightly underperform on rhythmic tasks. Notably, chroma features outperform MusicGen Decoder LMs on stationary harmonic tasks (notes, scales, and chords) but lag behind on dynamic harmonic tasks (chord progressions and intervals).

The aggregate handcrafted features perform comparably to MusicGen Decoder LMs. This suggests that harder music concept understanding benchmarks should address concepts latent in foundation models but not easily encoded in handcrafted features. These harder benchmarks may include entangled concepts, such as probing for both chord progression type and tempo in a chord progression samples of varying tempos. Probing for compositional tasks could reveal differences in concept encoding between model representations and handcrafted features.

While these models have shown that they encode representations for music theory in audio, some state-of-the-art music foundation models primarily rely on text controls. In a future study, we plan to probe for language representations in music generation models to improve text-controllable methods.

## 6. CONCLUSION

We introduce SynTheory, a synthetic dataset of music theory concepts, that is concept-isolated, copyright-free, and scalable. We use SynTheory to evaluate the degree to which music theory concepts are encoded in state-of-the-art music generative models. Our experiments suggest that music theory concepts are indeed discernible within the latent representations of these models. We believe this will allow us to understand how to isolate and manipulate such concepts, which advances towards controllable generation and music theory benchmarks. We encourage the community to build more challenging probing datasets with our framework to further understand the relationship between symbolic and audio-based music generation.

## 7. ETHICS STATEMENT

Our work aims to understand if music generation models encode music theory concepts in their internal representations. Our dataset may be used to assess music generation models and may be applied towards fine-grained, music-theory based controllable generation.

Our custom dataset, SynTheory, is based on elementary Western music theory concepts and is generated programmatically. The data does not infringe copyright of musical writers or performers.

## 8. ACKNOWLEDGEMENTS

We would like to thank Professor Cheng-Zhi Anna Huang, Professor Daniel Ritchie, Professor David Bau, Professor Jacob Andreas, Tian Yun, Nate Gillman, and Calvin Luo for their fruitful discussions and feedback towards this work. This project is partially supported by Samsung. We would also like to thank the Center for Computation and Visualization at Brown University for their computational resources towards this project. Finally, we greatly appreciate the insightful questions and thoughtful feedback from the reviewers.